\documentclass[final,3p,times,twocolumn]{elsarticle}
\usepackage{amsmath}
\usepackage{amssymb}
\usepackage{lineno}

\journal{Nucl. Intr. and Meth. A}

\begin{document}
\begin{frontmatter}
\title{Calibration of the TWIST high-precision drift chambers}

\author[triumf]{A.~Grossheim\fnref{htselsevier}}
\author[triumf]{J.~Hu\fnref{jlnow}}
\author[triumf]{A.~Olin}
\fntext[jlnow]{Now at AECL, Mississauga, ON, Canada, L5K 1B2}
\address[triumf]{TRIUMF, Vancouver, BC, Canada, V6T 2A3}
\fntext[htselsevier]{Corresponding author. Email: alexander.grossheim@triumf.ca}

\begin{abstract}
  A method for the precise measurement of drift times for the
  high-precision drift chambers used in the TWIST detector is
  described. It is based on the iterative correction of the space-time
  relationships by the time residuals of the track fit, resulting in
  a measurement of the effective drift times. The corrected drift time
  maps are parametrised individually for each chamber using spline
  functions.
  Biases introduced by
  the reconstruction itself are taken into account as well, making it
  necessary to apply the procedure to both data and simulation.
  
  The described calibration is shown to improve the
  reconstruction performance and to extend significantly
  the physics reach of the experiment.
\end{abstract}


\end{frontmatter}

\newcommand{\mevc}{MeV/$c$}
\newcommand{\kevc}{keV/$c$}
\newcommand{\um}{\mbox{$\mu$m}}
\newcommand{\app}{$\approx$}

\section{Introduction}
The TRIUMF Weak Interaction Symmetry Test (TWIST) experiment measures
momentum and angle distributions of positrons from the decay of
highly polarised positive muons to obtain an accurate measurement of
the decay parameters.

\begin{figure}[!hbt]
    \includegraphics[width=3.0in]{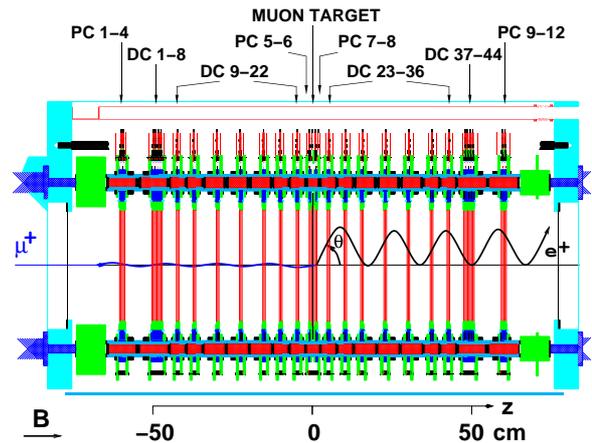}
    \caption{(Colour on-line.)  Side view of the
      TWIST detector.
      It consists of two symmetric stacks of drift
      and proportional chambers (DCs and PCs) surrounding the muon
      stopping target and is immersed in a 2 T magnetic field.
      A typical event is shown with the helical track of the decay
      positron.
    }
    \label{fig:stack}
\end{figure}

High-precision planar drift chambers are employed to reconstruct the
positron tracks in the momentum and polar angle range of $15 \lesssim p
\lesssim 55$ \mevc~ and $0.5 \lesssim |\cos\theta| \lesssim 1.0$,
respectively. The chambers are contained in a solenoidal spectrometer
with a 2 T magnetic field, and arranged symmetrically
around a central target foil that stops the low energy muon
beam (Fig. \ref{fig:stack}). Following the decay of the muon, the
positron is tracked in the chambers and a high statistics decay
distribution is acquired. The decay parameters are then extracted by
comparing the measured spectrum with a simulation.

TWIST's physics goal is an improvement in the accuracy of the decay
parameters by one order of magnitude over previous
measurements. Intermediate results, obtained from data taken before
2005, have been published \cite{rhopaper05,delta05,blair,rob2008}. To
further reduce the uncertainties for the analysis of the final data
sets acquired in Oct-Dec 2006 and May-Aug 2007, a detailed
understanding of the response of 
the drift chambers, in particular the space-time relationships (STRs),
was required. 

While the determination of drift time maps is a common problem, the
chamber setup and the precision requirements for TWIST made it
necessary to calculate accurate STRs directly from data. Consequently,
a method was devised to measure the drift times individually for each
chamber, taking into account construction inaccuracies, environmental 
parameters and interplay with the track reconstruction algorithms.

\section{TWIST Drift Chambers}
The 44 planar drift chambers (DCs) are the main tracking devices of
the TWIST spectrometer. A more complete description of the design and
construction of the DCs than is given here can be found in
\cite{TWIST_DCs:2005}.

\subsection{Chamber Design}
A primary design requirement for the DCs was to minimise their
total thickness, thereby reducing the amount of material that particles
have to cross when passing through the detector. This minimises 
the uncertainties from the calculation of energy loss 
and multiple scattering. At the same time, the chambers
had to obtain single-hit reconstruction efficiencies of above 99\% and
a spatial resolution of significantly better than 100 \um~ to achieve
the physics goals of the experiment. 

\begin{figure}[!hbtc]
  \begin{center}
    \includegraphics[width=3.0in]{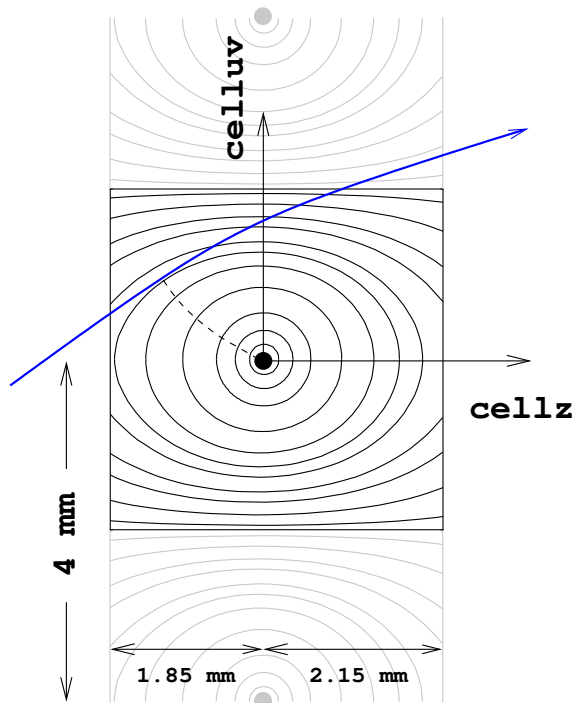}
  \end{center}
    \caption{(Colour on-line.) 
      Schematic layout of one drift cell and a projected particle trajectory.
      The cell coordinate system
      is defined with the origin at the wire and $cellz$ and $celluv$
      ranging from -1.85 to 2.15 mm, and -2.00 to 2.00 mm, respectively.
      The isochrones for drift times of
      2, 5, 10, 25, 50, 75, 100, 150, 200, 300, 500 ns
      are shown.
      As can be seen, the point of the charge deposit that reaches the
      wire first depends on the entry point and angle of the
      track. Its approximate path to the wire is indicated.
    }
    \label{fig:chamber}
\end{figure}

An individual chamber is composed of two parallel cathode foil discs of 320
mm radius, separated by 4 mm, and a wire plane with 80 parallel sense
wires stretched across the disc at
a pitch of 4 mm.
The foils are located at 1.85 and 2.15 mm distance,
defining asymmetric half-cells.
\footnote{The original design called for equal foil distances of 2
  mm. However, a systematic shift of the foil position resulted from
  the construction process. The wire planes are not affected.
}
No field wires are used.
The schematic layout of one drift cell is shown in
Fig. \ref{fig:chamber}. A drift cell is defined as the projection of
the drift space along all wires of a particular plane.
Sense wires are 15 \um~ diameter gold-plated
tungsten/rhenium with
the initial distance between wires accurate to approximately 3.3 \um~ and the
absolute wire positions known to better than 20 \um. The
foils are aluminised Mylar with a nominal thickness of 6.35 \um.
As drift gas
dimethyl ether (DME) at atmospheric pressure is used. In addition to
having a low mean atomic number, DME has a small Lorentz angle, a good
cluster yield (\app 30 cm$^{-1}$) and a relatively low drift
velocity, with the maximum drift times being below 1 $\mu$s for the
given geometry.

\subsection{Stack Assembly}
As shown in Fig. \ref{fig:stack} most DCs are assembled into
modules of two chambers, with the central cathode foil shared, and the
outer 
foils serving as gas containment windows. The two wire planes of such
a module are rotated by $\pm$ 45 degrees with the detector axis to 
reduce gravitational sag. The
wire orientations then define the $u/v/z$ coordinate system, with $z$
being the detector axis. Fourteen such modules are symmetrically
arranged around the target, with irregular spacing to reduce
ambiguities in the reconstruction. At the very upstream and downstream
these modules are complemented by two stacks of eight DCs that have
all inner foils shared. Upstream and downstream DCs are flipped with
respect to each other so that the smaller half-cell is always oriented
toward the target. In addition, proportional chambers (PCs) surround
the DCs at the very upstream and downstream end of the detector, and
around the target. They provide trigger and timing information and are
not used for the track fit.

The precise $z$-positioning of all modules is implemented with ceramic
spacers that have an extremely low thermal expansion coefficient. They
stabilise the distances between two wire planes to an accuracy of
better than 5 \um. The positioning of the foils is accurate to about
100 \um~ with respect to the wire planes and is expected to vary from
chamber to chamber. The space between the individual chamber modules
is filled with a (97:3) mixture of helium\footnote{Helium is used in
  order to reduce the amount of material beam muons and decay
  electrons have to traverse. The use of air would significantly increase
  the energy loss and associated systematic uncertainties.} and
nitrogen at atmospheric pressure.

\subsection{Operation}
The pressures of the DME and He/N input lines are carefully
adjusted and monitored to avoid differential pressure that would bulge
the cathode foils. In particular, fast temperature changes of the
detector environment have to be avoided to allow the control systems
to adapt. The dynamic bulging of
the foils was not bigger than 35 \um~at the centre and does not
constitute a considerable source of reconstruction uncertainties.

With unavoidable variations in the atmospheric pressure the density of
the drift gas varies as well. This effect is accounted for in the
determination and usage of the STRs (see below).

The chambers operate at a voltage of 1950 V, providing efficiencies of
above 99\% with no significant risk of sparking. Signals on the wires
of all chambers pass through preamps and post-amplifier-discriminators
before being read out by LeCroy 1877 multi-hit TDCs with 0.5 ns
time resolution.

\subsection{Drift Times}
Initially, the drift properties of the chambers have been studied
using the GARFIELD \cite{garfield} program. GARFIELD calculates the
drift times expected for a given setup, including the specifications
of the gas, and the electric and magnetic fields. While an accurate
drift time map can be obtained this way, there are inevitable
differences with the real chambers, arising from construction inaccuracies,
local variations of the electric field, and temperature and pressure
variations.
The impact of each of these effects on the drift times can be
quantified using GARFIELD with modified chamber parameters, however a
comparison with the real drift times of all 44 chambers cannot be
obtained this way. This lack of knowledge, although difficult to
quantify, could represent significant systematic uncertainties in the
final TWIST results.

Drift times can be determined
by using calculations, or data from test beams or dedicated 
calibration runs. These calibrations are often simplified by a static
and well determined geometry of the chambers, or a symmetric setup
where the drift time is a function of the drift distance only. An
additional simplification may arise from a limited topology
(curvature, angle) of the tracks traversing the sensitive
volume. Another source of STR biases is the simulation of the
ionisation rate of the drift gas which determines the separation of
primary ionisation clusters. These are simulated using a cluster
spacing matched to the shape of the time spectrum from tracks passing
very close to the sense wire. Lastly, the algorithm used to reconstruct
hits (i.e. assigning a spatial coordinate to a drift time) and track
fitting may introduce biases. If any of those biases are different for
data and simulation, or for calibration and physics data,
uncertainties on the physics results arise that are difficult to 
quantify.

To reach the precision needed to achieve TWIST's physics goals all of
the above-mentioned concerns had to be addressed: small variations in
the geometry, hence the electric field, of the 44 chambers are
expected, drift cells are asymmetric and a two-dimensional
parametrisation of the drift times is required, temperature and
pressure gradients may be present, and tracks cross the drift volume
at a large variety of angles and curvatures. In addition,
it is important that
approximations that are made in the TWIST track reconstruction impact
data and simulation in the same way.

\section{Track Reconstruction}
A certain interplay between features of the reconstruction, in
particular the track fit, and STRs can be expected. While the
reconstruction was \emph{not} changed for the application of this
calibration, it is important to understand the algorithms that are
employed.

\subsection{Pattern Recognition}
To reconstruct a track, the DC hits ---signal times on individual
wires--- are first grouped based on timing information from the PCs. A
combinatorial, geometric pattern recognition is performed on the hits
in such a time window to assign groups of hits to a potential track
candidate. For each track candidate an initial helix fit is performed
by approximating the position of hits by the crossing
coordinates of a pair of hit $u$ and $v$ wires in a module along with the
module's $z$ position. This gives the starting values for the
drift-time fit. 

\subsection{Helix Fitting}
\label{sec:hefitter}
The drift-time fit  is the simultaneous least-squares minimisation of
the sum of 
drift distance residuals of all points and the scattering angles
between them. The residuals are calculated by
approximating the track's trajectory through a drift cell by a
parabola, and finding the point of minimum drift time along that
track. This is the expected signal time for this track, and the
difference to the measured time is the time residual $t_{\text{res}}$. It can
be translated into a drift distance residual by swimming orthogonal to
the isochrones (see Fig. \ref{fig:chamber}) to find the distance
corresponding to a time difference 
of $t_{\text{res}}$. In practise, the drift distance is only varied radially.
This
approximation is made for performance reasons, but introduces biases
depending on the position of the hit in the cell, and therefore the
track angle. Individual drift distance residuals are weighted by the
chamber resolution at the point where the hit is placed and then added
to the $\chi^2$ function to be minimised.

The shape of the resolution function used to weight residuals is given
by the width of the time residual distributions for most parts of the
drift cell.
Close to the wire, an additional mechanism is employed to reduce  the
weight of points for which the so-called left-right (L-R) ambiguity is
not resolved yet. In those cases, during the first few fit iterations
the hit could still be placed on either side of the wire, leaving two
realistic drift time minima. 

The presence of multiple scattering is taken into account by
allowing kinks \cite{lutz88:kinks} between track segments. Seven such
segments are defined between the centres of DC modules. Kinks are
added to
the $\chi^2$ function, weighted by the expected scattering angle for
the amount of material crossed. This reduces angle-dependent biases
such as tracks being split into segments by the reconstruction due to
a large-angle scatter. In addition, the helix fit is not a simple fit to a
geometrical helix. In order to obtain better resolution and minimise
biases, the average energy loss of the particle and hence the
diminishing radius in the magnetic field is taken into account. The
fit result, typically after 10-15 iterations, is the particle's time,
position and momentum vector at the chamber closest to the target that
has contributed a hit.

\subsection{Usage of Drift Times}
Drift times are provided to the fitter in the form of one
$t_{\text{drift}}(cellz,celluv)$ STR table per DC plane, with a uniform
granularity of 10 \um~ in each dimension.
Since the distance between the wires does not
significantly deviate from the design values, the electric field and
the drift times are defined
symmetrically along the $celluv$ axis. The same symmetry can not be
applied along the $cellz$ axis as the distance between wires and
foils is not symmetric, and is associated with a rather large
uncertainty in the foil position.

During the actual fit, the STRs are used both to find the
minimum drift time along a trajectory,
and also to convert the drift time residual into a drift 
distance residual. These operations are executed several hundred times
for each track and must be optimised to reduce the computing time
needed. Thus, the STRs are not used directly as a lookup
table, but rather parametrised and converted into formats suited to
accelerate these two particular search patterns.

\section{Creation of new STRs}
Calibrated STRs are calculated in an iterative procedure starting from STRs
calculated by GARFIELD. At each step the time residuals are accumulated
as a function of the hit position within the cell. These are applied
as a correction to the input STRs in order to calculate the drift time 
table for the next iteration. For data, this is done individually for
each chamber while for MC all chambers are treated as one.

The procedure to create new STRs is described here in some detail as
implementing these steps correctly has proven to be challenging, 
but critical to the success.

\subsection{Determination of Time Residuals}
For the analysis of time residuals the drift cell is divided into 1044
bins (sub-cells) ranging from a size of (100 \um~ $\times$ 50 \um) near
the wire to (200 \um)$^2$ at the edges. Using the symmetry about the
$cellz$ axis, points with negative $celluv$ are mirrored, leaving
522 sub-cells for the range of -1.85 mm $< cellz <$ 2.15 mm and 0.0 mm
$< |celluv| <$ 2.0 mm. For each of those sub-cells, time residuals are
stored in a histogram with 1 ns binning. These distributions are
approximately Gaussian and a two-stage fit is used to extract the peak.
To populate all sub-cells sufficiently, approximately
$300\times10^6$ events\footnote{The amount of raw data to be processed
  at each iteration is up to 0.8 TB, requiring a few days of data
  taking, and approximately 0.5
  CPU-years of computing time for the reconstruction. With the
  resources used for this 
  calibration the latter typically took 2 days to complete.}
are processed, accepting all successfully fit tracks within a momentum
range of 10-55 \mevc~ and an angular range of 0.15 $<|\cos\theta|<$
0.95.

The calibration is performed off-line, and can use any decay data
destined for physics analysis. As such, data from various data-taking
periods are investigated to establish that STRs only depend on gas
density, and are the same for all data after density variations are
accounted for (see below).

\begin{figure}[!hbt]
    \includegraphics[width=3.0in]{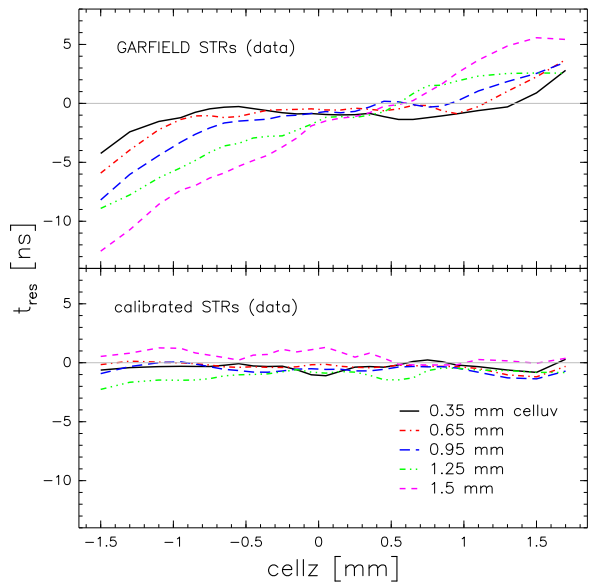}
    \caption{(Colour on-line.)
      Time residuals for an individual chamber (DC 2)
      before the calibration using GARFIELD STRs (top), and with the
      final calibrated
      STRs (bottom). The 
      individual graphs are for various $celluv$ slices
      (c.f. Fig. \ref{fig:chamber}) as indicated.
    }
    \label{fig:tres}
\end{figure}

For most chambers the initial time residuals show a variety of
distinct patterns, and can reach up to 15 ns in the corners of the
drift cell (see Fig. \ref{fig:tres} for a typical example). These time
residuals quantify the difference between the real drift times and the
ones calculated by GARFIELD, and represent the correction that needs
to be made.

\subsection{Fitting new STRs}
Various methods have been explored for the parametrisation of the new
STRs and the application of the correction defined by the time
residuals. These are not straightforward, as the corrections are not
measured in the same fine binning that is needed for the STR table,
and the final STRs obviously need to satisfy certain smoothness and
continuity requirements.

Under these conditions, STRs are most suitably represented by spline
functions, and not by functional forms derived from theoretical
considerations on the shape of the field. This is understandable
considering that the helix fit algorithm introduces a small
non-physical bias on its own, and given that the electric field itself
might be deformed in a variety of shapes, resulting from deformations
of the foils, for instance a tilt or bulge. 

The best method for the subtraction of the time residuals was found to
be the following algorithm. First, the fine-grained input STRs are
down-sampled to the coarse binning of the time residuals so that the
latter can simply be subtracted bin by bin. The resulting STR map is
mirrored about the $cellz$ axis and fitted with a two-dimensional
third order B-spline.\footnote{The SciPy (FITPACK) software library is
  used for the spline fitting.}
To satisfy the boundary condition of
$t_{\text{drift}}=0$ at the wire, a region of 250 \um~ radius around the wire
is blanked for the fit, and replaced with one individual highly
weighted point at the wire and zero drift time. The resulting spline
function is then used to calculate the drift time table for the 10
\um~ grid used by the helix fitter.

The smoothness parameter for the spline fit is adapted individually
for each plane in a compromise to obtain a close representation of the
data points without introducing oscillations between points to which spline
fits are prone. Other parameters, such as the number and position of
knots of the spline are left free.

Before accepting a set of new STRs as basis for the next iteration, a
test run is performed on a small amount of data. Several criteria are
used to estimate whether these STRs represent an improvement over the 
previous iteration. The most direct indicator is a comparison of the
distributions of the helix fit $\chi^2$: on average, the track fit
residuals should obviously decrease, and more tracks should be
reconstructed with a smaller $\chi^2$. 
In addition, the total number of successfully fit tracks should not
decrease, while the computing time spent on an individual track fit
should decrease.
If these requirements are not met, this particular set of STRs is
discarded and a new one is created, based on the same time residuals,
by variation of the spline fit parameters. This procedure can easily
require dozens of attempts, especially for data where the STRs for 44
planes need to be tuned individually, and no viable way of automation
could be found. 

\subsection{Drift Gas Density Correction}
Drift times are expected to vary with the density of the DME chamber
gas. The pressure in the chambers is set to follow the atmospheric
pressure, and together with inevitable changes in the environmental
temperature this causes variations of the density. After a few  
iterations of the residual analysis described above, these variations
manifest themselves in a time dependency of the drift time residuals. 
Time residual analyses of data from different time periods with
significant differences in atmospheric pressure are used to parametrize
the changes of drift times depending on the DME density.
It is found to be consistent with
predictions from GARFIELD and the maximal drift time variations are of
the order of a few percent and the same for all planes and all parts
of the drift cell.
Consequently, for both the reconstruction and STR
determination, input STRs are scaled according to the DME density
based on temperature and atmospheric pressure measured for any given
run, corresponding to a time frame of about 10 minutes.
Other significant time-dependent variations of drift times have not
been found. This is to be expected as long as the chambers remain
mechanically unmodified.

\begin{figure}[!hbt]
    \includegraphics[width=3.2in]{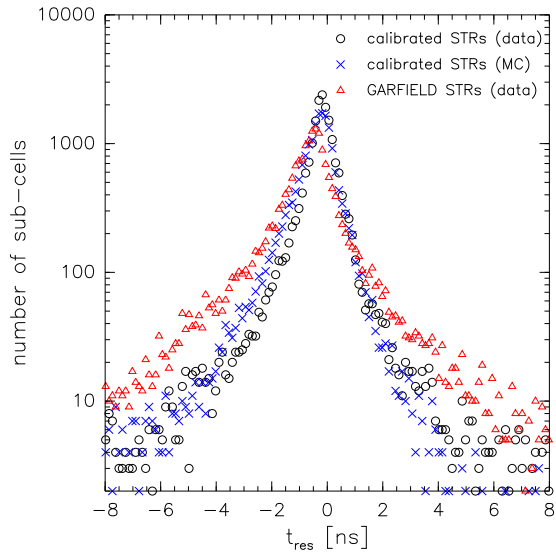}
    \caption{(Colour on-line.)
      Histograms of the time residuals of all sub-cells in the detector
      (44 planes times 522 sub-cells per plane). While there is still a
      slight difference between data and MC when calibrated STRs are used, a
      significant reduction of the tails are
      achieved. The latter are typically caused by asymmetries within
      the cell as shown in Fig. \ref{fig:tres} and give rise to
      reconstruction biases.
    }
    \label{fig:treshisto}
\end{figure}

\subsection{Convergence}
The procedure described above can be considered converged when no further
improvement of track fit $\chi^2$ can be achieved. For MC, where all
DC planes are the same, this typically requires 3-5 iterations.
As a crosscheck, a plane by plane analysis was performed on MC to confirm
that the algorithm itself does not introduce any significant
plane-dependent bias. 
For data, about twice as many iterations are needed as considerable
plane by plane variations need to be corrected.
The time residuals at the end of the procedure are mostly within $\pm$
1 ns, and no longer exhibit particular patterns that would
lead to biases in the reconstruction (see Figs. \ref{fig:tres},
\ref{fig:treshisto}). In 
spatial coordinates, the correction can be quantified as up to
40 \um~ in terms of shifts of individual isochrones.

On top of plane-dependent STRs, an analysis was performed to evaluate
whether there are significant differences of time residuals in
different regions of an individual DC plane. While minor differences
were found, they were considered too small to warrant major efforts to
correct. Moreover, unlike plane-to-plane differences, they should
average out since the muon decay spectrum is rotationally invariant.

\section{Resulting Improvements in the Analysis}
The overall goal of any calibration in TWIST is to improve the
consistency between data and MC. This is particularly important for
the momentum and angle dependency of reconstruction biases, resolution
and efficiencies. Such differences between data and MC would cause
first order biases in the measured decay spectrum parameters and
impact TWIST's physics reach. While the possibilities of directly
evaluating the above reconstruction benchmarks for data 
are very limited, some improvements to
the track reconstruction as well as on decay parameters can be
quantified as follows.

\subsection{Track Fitting}
\begin{figure}[!hbt]
    \includegraphics[width=3.0in]{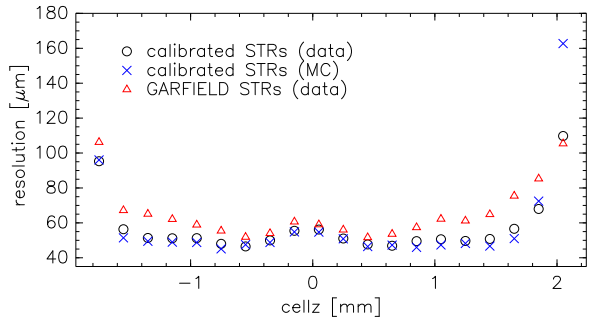}
    \caption{(Colour on-line.)
      Chamber resolution for a $celluv$ slice in the middle of the
      drift cell ($celluv$=0.95 mm).
      Using calibrated STRs for data allows the reconstruction to
      reach a similar resolution as for simulation. For the latter the
      reconstruction performance does generally not change
      significantly between  
      using GARFIELD or calibrated STRs and
      only one of them is shown.
      The resolution achieved is roughly consistent with the 30-50
      \um~ predicted by GARFIELD and confirmed in bench tests of
      individual chambers (c.f. Fig. 15 of
      Ref. \cite{TWIST_DCs:2005}).
    }
    \label{fig:cellreso}
\end{figure}

Some features of the helix fits, such as residuals, the average number
of points per track, the number of iterations, etc., can be used to
evaluate the quality of the reconstruction, and to characterise
how well the simulation represents the behaviour of the data. All of
these benchmarks indicate that an improvement in the absolute
performance as well as in the data-MC consistency is obtained.
As an example the chamber resolution is shown
in Fig. \ref{fig:cellreso}. It is derived from the width of the
distribution of drift distance residuals of the helix fit for all
planes. One can see that with the use of calibrated STRs the chamber
resolutions for both data and simulation match within 5 \um~ for most
parts of the cell, compared to a systematic difference of up to 20
\um~ before. With the hit 
position predominantly depending on the track angle, this reduces
angle-dependent data-MC differences and also proves that the combined
generation and reconstruction of hits is properly modeled in the MC.

The only test that directly allows a limited comparison of the momentum
reconstruction 
bias and resolution in data and simulation is the evaluation of the
kinematic edge of the spectrum at a momentum of \app 52.8 \mevc. In
the absence of physics beyond the Standard Model, its
theoretical shape and position are known, and any smearing and shift of the
reconstructed edge represents the momentum resolution and bias,
respectively. At a fixed momentum, the resolution scales with the
track radius and is proportional to $1/|\sin\theta|$. The value at
$\theta=90^\circ$ can be used for comparisons and it is found that for
both data and MC the momentum resolution is around $58\pm1$ \kevc~
while for previous 
analyses (e.g. \cite{rob2008}) it was typically 73 \kevc~ in data and 69
\kevc~ for MC. Similarly, calibrated STRs make the angular
dependence of the momentum bias for data to be 
identical to the one in MC. These improvements lead to a more reliable
energy scale calibration which is based on difference in the edge
position between data and MC and has to be accurate to the level of a
few \kevc.

\subsection{Muon Decay Parameters}
For most analyses completed prior to the development of the
calibration described  in this article, the dominant systematic
uncertainty arose from the lack of knowledge of the STRs
\cite{rhopaper05,delta05,rob2008}: While for MC the same GARFIELD STRs
were used for the generation \emph{and} analysis, data was
reconstructed with GARFIELD STRs that could not be validated
conclusively against the real drift times. 

The corresponding uncertainties were evaluated by the comparison of
drift time spectra in data and MC, and later more accurately by using
an early version of calibrated STRs. For the last pre-calibration
analysis \cite{rob2008}, the STRs were found to be the dominant source
of uncertainties and increased the total systematic uncertainties of
the experiment by up to 60\%.\footnote{Different decay parameters have
  a different 
  sensitivity to the uncertainties related to STRs, see
  \cite{blair,rob2008} for details.} 

For the final TWIST analysis using calibrated STRs, the related
uncertainties are estimated using the difference in the remaining time
residuals between data and MC. They are found to be at least a factor of
5 smaller than before, and contribute less than 7\% of the final
systematic errors in the decay parameters. These
significant improvements in the formerly leading systematic 
uncertainty are crucial to enable TWIST to reach its physics goals.

\section{Summary}
A method was devised to calibrate the space-time relationships of the
TWIST high-precision drift chambers. It is based on an iterative
procedure to correct drift time maps using the time residuals of the
track reconstruction fit. Space-time-relationships are then
parametrised by splines, individually for each of the planes in the
detector. Differences between individual planes, as well as variations
in operating conditions are corrected this way. The calibration is
applied to data and simulation, reducing systematic differences
such as reconstruction biases between both.

These improvements lead to a significant reduction of the related
systematic uncertainties in the measurement of muon decay parameters
that had limited previous analyses, and enable TWIST to extend its
physics reach.

\section*{Acknowledgements}
We wish to thank
C.A. Gagliardi,
A. Hillairet,
R.P. MacDonald,
R.E. Mischke, 
K. Olchanski, 
R. Poutissou, 
A. R. Rose, 
V. Selivanov
for sharing their expertise in valuable discussions. The support and
advice from all members of the TWIST collaboration, as well as staff
at TRIUMF and collaborating institutes is gratefully
acknowledged. This work was supported in part by the Natural Sciences
and Engineering Research Council of Canada. Computing resources for
the data analysis were provided by WestGrid.

\bibliographystyle{elsarticle-num}

\end{document}